\documentclass[12pt]{article}
\usepackage[T1]{fontenc}
\usepackage{a4,times}
\usepackage[colorlinks]{hyperref}
\newcommand {\lien}[1]{{\tt \href{#1}{#1}}}

\begin{document}

\title{A measure of similarity between scientific journals and of diversity of a list of publications }
% peut-etre faut-il mieux parler de scientific diversity??
%Note - mesurer l'interdiciplinarité / l'originalité }
\author{S. Cordier, Stephane.Cordier@univ-orleans.fr }
\date{ Version 1.0 - oct. 2012}
\maketitle

% plan de travail
% finir la redaction et faire une note sur Hal et arxiv
% envoyer a AERES (L. Dugard et al), MI (asch, massiot,verges) + gokalp + degond + AM Jolly
% envoyer a AMISC (demander suggestions) 
% envoyer a fortunato, newman, porter...
% ecrire aussi a ISI, researgate, HAL, google/scholar,  iop, springer  http://researcherid.com/
% citeseer : http://citeseer.ist.psu.edu/
% regarder centralité/ similarité
% http://fr.m.wikipedia.org/wiki/Diagramme_global_d%27interaction   
% serge bauin

% a envoyer a M. Langlais, P Vigneaux, D Talay

{\bf Abstract } : The aim of this note is to propose a definition of the scientific diversity and
corollarly, a measure of the ``interdisciplinarity'' of  collaborations. With respect to previous studies,
the proposed approach consists of 2 steps : first, the definition of similarity between journals and
second, these similarities are used to characterize the homogeneity (or, on the contrary
the diversity) of a publication list (that can be for one individual or a team).

\section{Introduction}

Interdisciplinarity is, nowadays, of interest  for several reasons and by lots of people and
institutions.  Let us just quote two recent initiatives in France : the creation of the "mission
interdisciplinaire" at CNRS ( \lien{http://www.cnrs.fr/mi/})  or the report of the AERES that proposes
interesting direction for evaluation of interdisplinary \cite{aeres} based on qualitive analysis.
% citer IUF ?
We do not intend to discuss the reasons of such interest and refer to \cite{wagner2011}
for a detailled and recent review about interdisciplinarity. \\
% on veut pas non plus savoir si les travaux interdisciplinaires sont "meilleurs" que d'autres?
% mais on pourrait regarder s'ils sont plus cités?

In this note, we propose a method for quantifying the interdisciplinarity only based
on bibliometric data, without any a priori classification of scientific domains and/or
arbitrary knowledge on their proximity. The obtained results should be compared with
existing classification  and analysed by scientists to validate (or not !) their meaningful interest.\\

The actual note is a very preliminary description of the idea and it has not been 
tested on bibliometric data. The author is not an expert in scientometrics and
do not have acces to large database that are necessary to test the approach.
Lot of studies have been done about co-authorship (see e.g. \cite{fortunato-2010} and 
the reference cited therein, at bottom of page 159).
However, %bibliographic researchs indicate that 
the actual approach (with two steps
as detailled after) has not yet been proposed up to my knowledge. In the actual version, this note 
is not aimed for publication and suggestions are warmly welcome 
in particular to be informed about previous works in the same spirit.\\

The goal is to define a measure of the interdisciplinarity within a publication list
(for one individual, team, laboratory, institution). Such quantitive information 
has to be complemented by a finer analysis for scientists to determine the corresponding
relevance of the scientific collaborations. We just try to propose an approach
to see its faisability and,  hopefully, to prove its capability to characterize interdisciplinary 
studies. \\

%{\bf Objectifs } : on cherche à trouver un moyen de quantifier la part de travaux interdisciplinaire 
%dans une liste de publications (d'un individu, d'une équipe, d'un laboratoire,?).
%Cette analyse quantitative est complémentaire d'une analyse qualitative fine qui seule
%pourra déterminer la pertinence des interactions, ce qui nécessite une expertise fine 
%et qui est naturellement 
%indispensable à un travail d'évaluation scientifique. L'objet de cette note est en quelque sorte d'une étude
%de faisabilité pour voir si les résultats d'une analyse automatique peut (ou non) donner
%des résultats pertinents. 

{\bf The proposed approach is based in two steps :  first, we define, from a bibliographical database,
a measure of the similarity between scientific journals based on co-authorship i.e.
the more 2 journals have co-authors, the closer they are (this will be made more precise later). 
It can be objected that we do not measure scientific ``proximity'' %(in a sense to be  define) 
but actual practices of publications.
The second step consists in using these similarities to characterize whether or not a
publication list is an scientifically ``homogeneous'' set.}\\

%Postulats / hypothèses : On part d'une liste la plus complète possible d'informations
%bibliographiques, contenant naturellement les productions des personnes dont on
%souhaite caractériser l'interdisciplinarité. On commence par construire une mesure
%de la proximité entre 2 journaux en se basant sur les publications des auteurs ayant
% des publications dans ces journaux. Ce faisant, on mesure en fait plutot les habitudes de 
% publications des différentes comunautés.   

It is believed that information on co-authorship is more reliable to evaluate pluridisciplinary 
collaborations than using citations. Indeed, it is rather common that a paper e.g. in mathematics
cites several articles in an application domain, to illustrate the origin of the scientific problem or
to justify the modelling choice done but the core of the paper can be entirely focused on
mathematical analysis. On the other hand, signing a paper with colleague mean (we hope so) a
mutual interest and work within the paper. 

Note also that publishing a paper in a so called pluridisciplinary journal (what is the  definition of such journals ?) 
does not mean that the article is itself the result of a collaboration between several scientific domain. 

It can be argued that the proposed approach measures more the originality of a set of publication,
that is the reason why it is called scientific diversity, since the similarity counts for the existing collaborations, 
even if they are already interdisciplinary.

Several web site indeed provide informations on scientific collaborations
such as {\it researgate, resaerchid, googlescholar, sciencewatch} (non exhaustive list)
and can be interested in providing new services and informations to their visitors
(see in the reminder for examples).

The proposed analysis can also be of interest for editors of scientific journal
(they may already have similar tools but they are not known by the author).
The method is presented in a algorithmical way in order to facilitate 
its implementation. Let repeat that experiment feedbacks are welcome.

%\section{Methodology}

\section{Data , notations}

The method relies on bibliographical data (the use of the largest possible database 
will provide the more relevant informations).  

% on pourrait discuter ici du bien fondé (relevance) de ce choix, i.e. de se fonder
% sur les publications dans les journaux a comite de lecture , sachant que le modele
% editorial est actuellement peut-etre en train de changer? mais cela nous entrainerait
% trop loin et ferait perdre le fil? a garder pour une version longue (ou une footnote ?)

Let us precise the notations.

The database consists in a list of articles, that will be denoted by an unique identifier
(that can be consider as in integer) using the letter $i$. 
Each article $i$ (where $i \in {1 \cdots N}$)  will be described by
\begin{itemize} 
  \item the journal of publication, denoted by an unique identifier, $j$.
  More precisely, $j(i)$ is the journal where article $i$ has been published.
  The list of journal is finite (even if its length increases with the creation of new journals
  each year). To fix the idea about 13000 journals are included in the Thomson-ISI database.
  \item $y(i)$ is the year of publication of the article $i$. 
  \item $K(i)$ is the list of (co-)authors of the article $i$. 
  The authors have to be identified i.e. each individual should have 
  an unique identifier, that can be represent by an integer. We will use
  $k$ for authors. Thus, $k \in K(i)$ means that the author $k$ is (one of)
  the author of article $i$
  \footnote{ This is the reason why I can not test the proposed approach on the 
  data of HAL french publication deposit.}
  \item $p(i)$ is the number of pages of the article. This is useful to differentiate
  short note and more detailed study although this can be discussed. The
  interest of a paper is, of course, not proportional to its length but it
  can be consider as an useful indicator, once renormalized for a given journal
  (or a given author). 
\end{itemize}   
  
%    {\bf Données}
%On suppose connus les informations suivantes, pour une liste d'article $a_i$ pour
%$i=1 \cdots N$ où $N$ est le nombre d'articles dans la base bibliographique 
%$$
%a_i= \{ j, y, K, n_k, p\}
%$$
%où $j$ est le (numéro du) journal (parmi une liste finie naturellement), 
%$y$ est l'année de publication/parution, $K$ est le nombre d'auteurs,
%$n_k$ est l'identifiant de l'auteur $k$ pour $k \in 1 \cdots K$
%$p$ est le nombre de pages.\\

In this  note, we will use capital letters to represent lists. 
$J$ is the (finite) list of all the journals. $I$ is the list of all the articles.
For example, we shall note $I(j)$ the list of articles  published 
by the journal $j$ or $I(j,k)$ the list of articles published by
author $k$ in the journal $j$ or $I(j,k,y)$ the list of articles published by
author $k$ in the journal $j$ within the year $y$.
Similarly $J(k,y)$ represents the list of journals where author $k$ published
in the year $y$.\\

We note with $N$ the cardinal of a set. E.g.
$N(I(k))$ is the total number of articles by author $k$. $P$ is the total number of pages, e.g.
$P(j,y)= \sum_{i' \in I(j,y)} p(i') $ is the total number of pages in the journal $j$ during year $y$.

%Le poids  d'un article sera réparti uniformément sur chaque auteur i.e. divisé par $K$ et on pourra 
%évenuellement le rendre proportionnel  au nombre de page i.e. $p_i/K$. \\
%De cela, on peut en déduire la liste des articles d'un journal $j$ donnée,
%d'une année $y$, d'un auteur $k$?. ou tout requête multiple (ex. le nombre
%de page publiée par l'auteur $k$ dans le journal $j$ pendant l'année $y$).\\

According to usual consideration in the mathematical science field (which is
the domain of the author), the weigth of a given article will be shared uniformly
between all the authors of the article. This point is naturally questionnable 
but claiming that the importance of a paper is proportional to its number of
authors as, e.g. in the computation of citations or impact factor 
 can also be  under discussion. 
Let us refer to \cite{derek} for a discussion  on the question of multiple authorship.\\

\section{Journal similarity}

Using these notations, we shall now define the similarity between journals
by considering, for all articles and of (co-)authors of this article, all other articles
by the same author.

More precisely, for all $i \in I$ and all $k \in K(i)$ and all $i' \in I(k)$, the 
similarity betwen journal $j(i)$ and $j(i')$ increases as follows 
\begin{equation}
\label{def_S}
S(j(i),j(i')) += \min( { p(i) \over N(K(i))} , {p(i') \over N(K(i'))} ),
\end{equation}
note that $S( j(i),j(i'))$ will be increased by the same value 
(when exchanging the role of $i$ an $i'$).
We propose to increment the similarity between the 2 journals by the
$\min$ value of the ``weight'' of the 2 articles (number of pages divided by
the number of authors) instead of using the arithmetic mean e.g. because
it is believed that  the scientific proximity is stronger if the 2 papers have 
the same weights (with the same arithmetic weigth).
%than if one is much more important than the other 
Other choices, like for example, a geometric mean ($\sqrt{ab}$) may give
better results. The only way to choose the right formulae will be to test 
several choices and compare the obtained similarity matrix (see below some
ideas to help in the choice or in the validation of the relevant definition of 
similarity).\\

% [8] de Solla Price, D. (1981). Multiple authorship. Science, 212:986.
% west-2012

One can discuss about the normalization of the page number  i.e. to divide the number of pages
with respect with the total number %(or the average number) 
of pages withinn the journal $j$.
In other words, we propose to replace $p(i)$ is the above equation by
$$
\tilde p(i) = p(i) /  P(j(i)). %({\mbox or }    p(i) N(I(j(i))) / P(J(i))
$$
These variants should be tested as soon as data are available.
It is obvious that the non-normalized choice will increase the impact of journals which produces a 
lot of paper and/or pages whereas the similarity should measure a proximity between journal that
should not be correlated to the "size"  of the journal. Therefore, the normalization using normalized 
by the total number of paper should be more pertinent.\\

%On peut se demander s'il faut (ou non) normaliser le nombre de pages i.e. diviser le nombre
%de page d'un article par le nombre total de pages pris en compte dans le journal. Si on ne 
%le fait pas, on donne naturellement un poids plus important aux revues qui éditent un plus
%grand nombre de pages, qui apparaitraitront donc plus proches des autres revues qu'une
%revue produisant moins de pages mais ayant la même politique éditoriale.

{\bf Remarks}\\

- By construction, $S(j,j) >1$ since $i \in I(k)$, $\forall k \in K(i)$
(or $S(j,j)> P(j)$ if we do not use the normalize version).
  The effective value will measure if the same authors are lot of their
  publications in the journal $j$.\\% (fidelity ? journal of a specific   community ).\\

- Similarity can be computed for a given year (or period) by taking into account 
only the article of the corresponding year (or period).\\

- It is clear that the matrix $S$ will be coarse 
%(with dominant diagonal ? to be proved? but it is not important in the reminder?)
and it will be necessary to take into account the second order co-authorship (i.e. the
co-author of co-authors, following the idea of the Erdos number ref?).
Let us define by summing the binary interactions as follows 
$$ 
S_2 (j,j') = \sum_{j"} S(j,j")S(j",j')
% autre idee mais plus douteuse ? \sum_{ k | S(j',J(k)) \not = 0} S(j,k). 
$$ 
By construction we see that $S_2=S^2$.
Then we can use $\tilde S = S + \theta S_2$ where $\theta$ is a constant than 
represents the relative weigth of secondary co-autorship. Once again, this should
be tested on a real / huge database (see below).\\

%- Remarque (distance) :
%Ensuite, on peut construire une pseudo-distance  en prenant 
%une fonction décroissante de $\tilde P$ :
%$$ d(j,j') = \varphi (\tilde P(j,j'))$$
%où $\phi(0) = + \infty $ et $\phi(x)=0$ pour $x>1$? mais cette 
%fonction ne vérifie pas nécessairement l'inégalité triangulaire. \\

%En effet, si deux revues $j$ et $j'$ sont très éloignées (car très peu de co-auteurs), 
%mais sont toutes deux proches d'une même revue $j"$ (qui pourrait être alors
%considérée comme "interdisciplinaire"), alors 
%$$ d(j,j') > d(j,j")+d(j',j").$$
%il est possible que ce problème disparaisse si on prend $\theta $ assez grand? 
%mais on n'a pas nécessairement besoin de cette "distance".\\

{\bf Validation phase }\\

At this stage, it will be  necessary to test if the proposed similarity fits with usually used classification by 
scientific domain.
More precisely, it will be interesting, using a given disciplinary classification, to verify wheater or not 
the averaged similarity inside a scientific domain is smaller (or not and to what extend) 
than the same average over all journals.
It can serve also to compute the average similarity between two choosen different domains by computing 
the average value of $S(j,j')$ for any $j$ in domain 1 and $j'$ in domain 2.  This will provide a similarity
matrix between scientific domains and it has to be analysed if it corresponds to usual classifications
 of scientific domains. 

Note that, when considering article from "multidisciplinary" journals (using an arbitrary list/classification), 
its citations are affected to a domain, according to the citation in the article
(see \lien{http://sciencewatch.com/about/met/classpapmultijour/}. \\

Other study like clustering can be developped using this similarity between journals \cite{newman-2010,GN-2002}.\\

It can also be checked if the "generalist" journals have a larger (average) similarity than the more specific one.
One can e.g. use the 22 so-called broad fields (see  \lien{http://www.in-cites.com/journal-list/} )
in  the ``Essential Science Indicators database'' of Thomson-ISI .\\

{\bf Possible Services - Utilities}\\

Once validated, this similarity between journals can be of interest for editors to evaluate the impact
of their editorial choice on the scientific positioning. For example, if an effort is done in order to 
encourage paper in a near domain (that corresponds to a given subset of journals), 
it can be observed that the average similarity of the journal
with the one of the given subset increases with time (by computing the averaged similarity restricted
to successive years). It can also help editors to see if a journal evolves for a larger specialization or on the contrary
is more and more multidisciplinary.

\section {Interdisciplinarity or scientific diversity index}
	
Let us now consider that the matrix of similarity  $S$ is known (and validated). \\

In this section, we will construct an index for any arbitrary list $L$ of publications
(that can be the one of a person, a team, a laboratory, an institution, a journal, 
an editor). 

Let us define the, so called scientif diversity index $SD$ of the list $L$ as the 
averaged similarity between journal in the list weighted with the respective 
weights of the article. In other words,  
\begin{equation}
\label{def_SD}
SD(L) = {1 \over N(L)^2} \sum_{i \in L} \sum_{i' \in L} S(j(i),j(i')) p(i)/N(K(i)).
\end{equation}
Note that the index is% ( this has to be check !) 
not related to the quantity of paper
(if one duplicates the list, the number of elements in the double sum is multiplied by 4 
but $N(L)$ is multiply by 2 and the value is unchanged). \\

The $SD$ index has not to be considered as an indicator of the quality of the articles in list $L$ , 
but on the contrary this is a qualitative indicator on this list of articles.
Note that this index is constructed using statistical / averaged bibliographical quantities.
It is therefore very questionnable to use it on a small list of articles and thus,
it is likely more suitable to characterize collective list of publications than
the one of individuals except for scientists with a sufficiently long publication 
list in order the result to be significative (for such scientists, it will be interesting
to see is their $SD$ is correlated with the number of articles ($I$ index i.e. 
$N(I(k)$ with our notation) or citations ($h$ index for example). 

They are lot of study that can be done using this index, which, again, does not measure
the "quality" or the "importance" or "impact" (in terms of influence on other scientists)
but only the relative diversity or variety or originality of a list of publications with 
respect to others. \\
  
%  De part la nature statistique, cette analyse nécessite un nombre important de
%publication et cela ne peut donc pas être utilisé facilement sur un seul individu
%(à moins qu'il soit très prolifique et/ou en fin de carrière) et cela donc est plutôt
%utile pour un ensemble de personne.

Such an indicator will only have to be used with lot of care and only relative 
comparison make sense (the exact value has no interest). 
For example, one can think to compute, for the list of a given author (denoted
by $I(k)$ with our notations), one can compare the value of $SD(I(k))$ with the
corresponding value for all co-author of $k$. \\

If such web service is implemented, it can be asked to the author if his/her proposed
ranking with respect to the one of his/her co-authors in term of "scientific diversity" 
seems to him/her relevant or not? This will be, in my opinion, a good way to evaluate if the proposed 
indicator gives information that fit with the general opinion. If some variant of the
above definition are proposed, it can be checked which definition better indicate
the scientific diversity.

\section{Central journal of a publication list}

One can also use the similarity matrix to define for any list $L$ the central journal 
by looking which journal in the list $J(L)$ maximise the similarity with 

\begin{equation}
\label{def_jc}
\bar j (L) = \{ j \in J(L)\; s.t. \;\; \sum_{i\in L} S(j,j(i))  {p(i) \over N(K(i))} = 
\max_{j' \in J(L)}  \sum_{i\in L} S(j',j(i)) { p(i) \over N(K(i))} \}.
\end{equation} 

In other words, if we interpret similarity as the inverse of a pseudo distance, the 
central journal is the one that will minimize the average distance with the
other in the list $J(L)$.  This may be related to the Fermat-Weber point of Fr\'echet mean.\\ 

%En fait, on pourrait interpréter cela comme la recherche d'un point dans un espace métrique 
%dont la distance moyenne (ou totale) est minimal de l'ensemble des autres ( l'ensemble étant fini, 
%la notion de moyenne et de total ne change que la normalisation de la somme).
%Question : est-ce que cette notion a un sens ? ( pour un intervalle continu, et la norme $p$, on
%retrouve le milieu de l'intervalle).\\

It may be interesting to rank the journals of the list by decreasing value of their 
averaged similarity with others (as defined above). It should give at the top of the list, the 
journals that correspond to the principal domain of the author or list  and, at the end,
 the journals that are scientifically far from his/her speciality.

As for the comparison of the scientific diversity (SD) of his/her co-authors,
one can think to ask, for such web service, if the ranking correspond to what
is usually admitted (using a pool, and eventually, comparing the results of 
other definition).

%Comment évaluer si les résultats obtenus sont pertinents ?
%--> Faire une enquête auprès des auteurs pour savoir s'il s'y reconnaissent ?
%On pourrait ainsi présenter, pour une personne donnée, la distance moyenne de
%chacun de ses co-auteurs , et leur demander si ce classement leur parait plutot
%satisfaisant ou plutot non relevant. 

One other possible service that can be useful for scientist is to propose some journals
in which they never have published but which are "close" in the sense that the 
similarity is high with their central journal (one can restrict the suggestions of paper
in the list of journals where their co-authors have published).
This can suggest to enlarge their list of journals and avoid scientific concentration.

%On pourra également classer les journaux suivant ce critère et éventuellement proposer
%à un auteur donné, des suggestions de journaux, dans lequel il n'a jamais publié mais
%où ses co-auteurs publient?. en les classant des plus proches au plus éloigné.. ou encore 
%des journaux très proches dans lequel il n'a jamais publié (ni ses co-auteurs directs).\\

It is also possible to give information about the evolution of their scientific diversity over years.
Note that this definition of a ``central journal'' is only an example of the use of the 
similarity index between journals and lot of other concepts can be proposed using tools of
graph theory, network analysis, clustering...\\

%On peut également faire l'analyse pour une période donnée et voir, pour un journal
%ou pour un (groupe d')individu si sa tendance est plutôt à la focalisation ou à l'ouverture
%thématique. 

{\bf Once again, feedbacks are welcome !}\\

{\bf Acknowledgment : }
The author would like to thank, by chronological order,  
S. Mancini (U. Orléans), 
L. Cappelli (CCSD, Lyon), 
V. Miele (CNRS, Lyon)

\newpage

% journal envisagé ? bibliometrics http://eu.wiley.com/WileyCDA/Section/id-380979.html

% Watts, D.J.; Strogatz, S.H. (1998). "Collective dynamics of 'small-world' networks.". 
% Nature 393 (6684): 409?10.  doi:10.1038/30918

% ref page 88 de l'article fortunato-2010
%http://www.itrmanager.com/tribune/135037/importance-pensee-systemique-quantitative-big-data.html

\bibliographystyle{plain}

\begin{thebibliography}{}

\end{thebibliography}


\begin{thebibliography}{1}

\bibitem{aeres} Critères d'évaluation des entités de recherche : Le référentiel de l'AERES, partie III, AERES, 2012.
see : http://tinyurl.com/9zzjwn2

 \bibitem{wagner2011} 
 % citation dans le rapport de l'AERES
C. S. Wagner, J. D. Roessner, K. Bobb, J. Thompson Klein, K. W. Boyack, J. Keyton, I. Rafols, K. Borner,
Approaches to understanding and measuring interdisciplinary scientific research : A review of the literature
Journal of Informetrics (2011). 

\bibitem{fortunato-2010}
S. Fortunato,  Community detection in graphs,	Physics Reports 486, 75-174 (2010).
	
%porter-2009
%Porter, A. L., & Rafols, I. (2009). Is science becoming more interdisciplinary? Measuring and mapping six research fields over time. Scientometrics, 81, 719?745

%Leydesdorff, L., & Rafols, I. (2009). A global map of science based on the ISI subject categories. Journal of the American Society for Information Science and Technology, 60, 348?362.

% Voir également European Science Foundation, Member Organisation Forum (2011), European Peer Review Guide. 
%Integrating policies and practices into coherent procedures http://www.esf.org/activities/mo-fora/peer-review.html.

\bibitem{derek} De Solla Price, D. (1981). Multiple authorship. Science, 212:986.

\bibitem{newman-2010} M. E. J. Newman, Networks: An Introduction,  Oxford University Press (2010). 



\bibitem{GN-2002} Girvan, M., and M. E. J. Newman, Proc. Natl. Acad. Sci. USA 99(12), 7821. (2002)

%Danon, L., A. D?az-Guilera, and A. Arenas, 2006, J. Stat. Mech. 11, 10.
%Donetti, L., and M. A. Munoz, 2004, J. Stat. Mech. P10012.
%Duch, J., and A. Arenas, 2005, Phys. Rev. E 72(2), 027104.
%Farkas, I., D. Abel, G. Palla, and T. Vicsek, 2007, New J. Phys. 9, 180.
%Lehmann, S., and L. K. Hansen, 2007, Eur. Phys. J. B 60,83.
%Nepusz,T.,A.Petroczi,L.Negyessy,and F. Bazso,2008, Phys. Rev. E 77(1), 016107
%Newman, M. E. J., 2004b, Phys. Rev. E 69(6), 066133.
%Newman, M. E. J., 2006a, Phys. Rev. E 74(3), 036104.
%* Palla, G., A.-L. Barabasi, and T. Vicsek, 2007, Nature 446, 664.

%++  Palla, G., I. Derenyi, I. Farkas, and T. Vicsek, 2005, Nature 435, 814.

%* Pujol, J. M., J. Bejar, and J. Delgado, 2006, Phys. Rev. E 74(1), 016107.
%Radicchi, F., C. Castellano, F. Cecconi, V. Loreto, and
%D. Parisi, 2004, Proc. Natl. Acad. Sci. USA 101, 2658.
%Reichardt, J., and S. Bornholdt, 2006a, Phys. Rev. E 74(1), 016110.
%Vragovic, I., and E. Louis, 2006, Phys. Rev. E 74(1), 016105.

\end{thebibliography}

\end{document}